# A Codon Frequency Obfuscation Heuristic for Raw Genomic Data Privacy

Kato Mivule
Computer Science Department
Bowie State University
kmivule@gmail.com

## Abstract

*Genomic data provides clinical researchers with vast opportunities to study various patient ailments. Yet the same data contains revealing information, some of which a patient might want to remain concealed. The question then arises: how can an entity transact in full DNA data while concealing certain sensitive pieces of information in the genome sequence, and maintain DNA data utility? As a response to this question, we propose a codon frequency obfuscation heuristic, in which a redistribution of codon frequency values with highly expressed genes is done in the same amino acid group, generating an obfuscated DNA sequence. Our preliminary results show that it might be possible to publish an obfuscated DNA sequence with a desired level of similarity (utility) to the original DNA sequence.*

## 1. Introduction

Genomic data grants clinical researchers with considerable prospects to study various patient sicknesses. Nevertheless, genomic data includes a lot of revealing information, some of which a patient might want to remain concealed. The question from such a patient controlled data retrieval scenario becomes: how can an entity transact in full DNA sequence data while concealing certain sensitive pieces of information in the genome sequence, and yet maintaining DNA data utility (usability)? In this paper, we respond to this question by putting forward a heuristic in which the patients might have a say as to what part of their genomic information they wish to remain concealed. The rest of the paper is organized as follows. In Section 2, a background definition of bioinformatics and data privacy terms used in this paper is given. In Section 3, a review of related work on the subject of DNA privacy is done. In Section 4, the proposed heuristic and methodology is outlined in detail. In Section 5, the experiment and results are discussed, and lastly in Section 6, the conclusion and future works is given.

## 2. Background

*DNA* stands for deoxyribonucleic acid, a chemical substance that comprises functionality information for the make-up and growth of an organism [1]. *DNA organization* is an organized collection of four building blocks called nucleotides bases, namely adenine (A), thymine (T), guanine (G), and cytosine (C), in which the adenine (A) pairs with the thymine (T), and guanine (G) with cytosine(C); forming a twisted ladder-like shaped strand, referred to as a double helix [2]. *Genomics* is the study of the genome (DNA) sequence structure, the chemical components that make up the DNA structure, including the chemical elements that hold DNA components together [3]. *DNA sequencing* refers to the process of mapping out the nucleotide sequence of DNA [4]. *Codon* is a sequence of three subsequent nucleotides in the genome that map to a specific amino acid in a protein; a codon could also represent a triplet, which is the start and stop points in the DNA sequence [5]. *Codon frequency table* or codon usage table is the statistical representation of each codon existence in the DNA sequence [6].

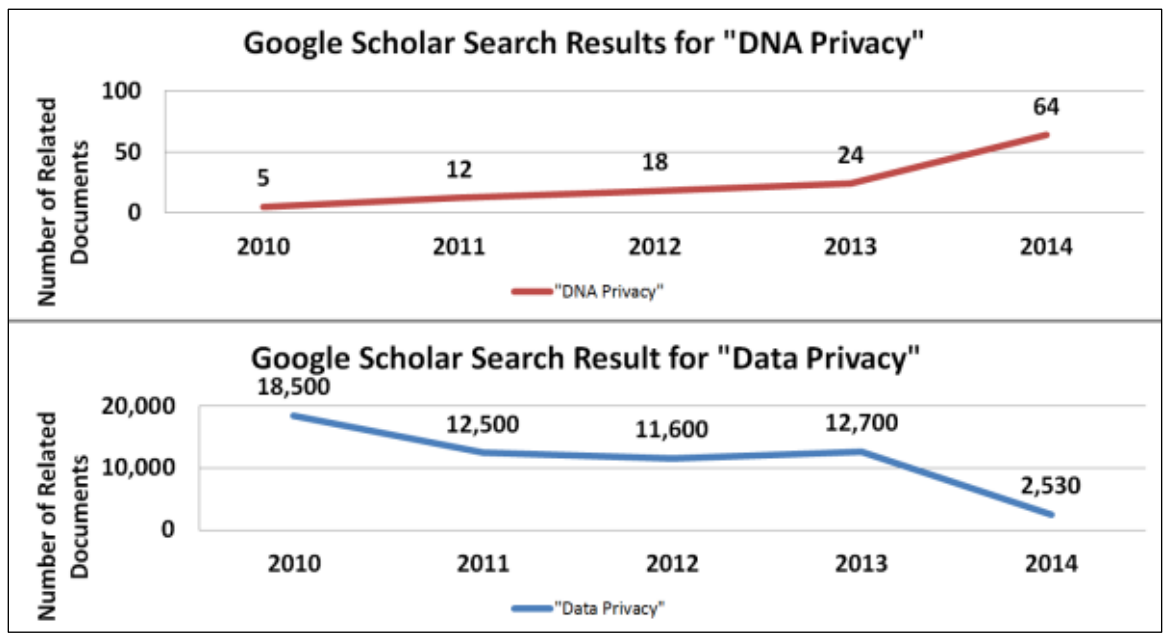

**Figure 1:** Google Scholar search results for "DNA Privacy"

*Data obfuscation* is a data privacy technique in which data is distorted, made imprecise, and indistinguishable so as to conceal information [7]. *Data shuffling* is a data privacy technique first proposed by Dalenius and Reiss (1978) as a data swapping methodology for categorical data [8], and later extended by Muralidhar and Sarathy (2003) for numerical data; shuffling is a data privacy procedure in which sensitive data values $v_i, \ldots v_n$, within an attribute $A_i$ are exchanged between tuples $t_i, \ldots t_n$ in the same attribute $A_i$; for instance, a sensitive value $v_i$ belonging to tuple $t_i$ is assigned to tuple $t_n$ and the value $v_n$ belonging to tuple $t_n$ is assigned to tuple $t_i$ in the same attribute $A_i$ [9].

## 3. Related work

In this section, we take a look at some work being done in the Genomics data privacy domain. Research in Genomics data privacy is relatively still in the early stages and not so much work and literature exists on the subject. For instance, as illustrated in Figure 1., a Google Scholar search results only returned 24 documents related to the "DNA Privacy" search term for 2013 compared to 12,700 related documents to the search term "Data Privacy", for 2013[10]. However, Genomic data privacy researchers have largely focused on using cryptographic techniques in granting confidentiality to DNA data sets. Malin and Sweeney (2004), proposed employing re-identification techniques in evaluating and modeling of anonymity protection systems for genomic data; the goal was to prevent any information leakage between genomic data and named individuals in public records [11]. Malin and Sweeney (2004)'s model focused on engineering a secure and confidential data privacy transaction model for genomics, without perturbing the DNA sequence structure. However, Malin (2005), proposed one of the first genomic data privacy models that involved the actual manipulation of the DNA sequence to enhance confidentiality. Malin (2005)'s model employed k-anonymity with generalization by ensuring that in a collection of DNA sequence data, each generalized DNA record would appear at least k > 1 times, to satisfy k-anonymity requirements [12]. Additionally, Malin (2005), provided a holistic evaluation of the current state of genomic data privacy storage and transaction systems then, by outlining the short comings and providing recommendations [13]. In this paper, we take a similar approach by proposing a genomic data privacy solution that takes into consideration both the bio-bank architecture and the confidential transaction of the individual's DNA sequence. Other works have centered on employing genomics as a tools for cryptography [14][15]. For instance Heider and Barnekow (2007), proposed using DNA cryptographic and steganographic algorithms for watermarks to detect any unauthorized use of genetically modified organisms protected by patents [16]. Additionally, Kantarcioglu, Jiang, Liu, and Malin (2008) proposed a cryptographic model in which aggregated data mining, such as, frequency counts, could be done securely on genomic sequences without revealing the original encrypted raw DNA sequence [17]. On the other hand, El Emam (2011), in a review of de-identification technics for genomics data, observed that electronic records are increasingly being linked to DNA bio-banks, used by clinicians for genomic research, and that challenges and risks still remained despite de-identification techniques for confidentiality [15]. In this paper, we argue for a holistic approach that considers privacy modeling at both the bio-bank storage level and the data transaction level between the patient, doctor, and clinical researchers. Ayday, Raisaro, and Hubaux (2013) proposed a holistic model for genomics confidentiality in which data privacy enhancing principles, using homomorphic encryption, were employed in modeling a secure bio-bank system, DNA data transaction process, and the confidentiality of the DNA sequence [18]. Furthermore, Ayday, Raisaro, Hengartner, Molyneaux, and Hubaux (2014) proposed a privacy preserving system to grant confidentiality using encryption on short reads of raw genomic data [19]. In their model, Ayday et al.,(2014), suggested storing short reads in encrypted form at a bio-bank and then allow the stakeholders to securely retrieve only needed short reads [19].

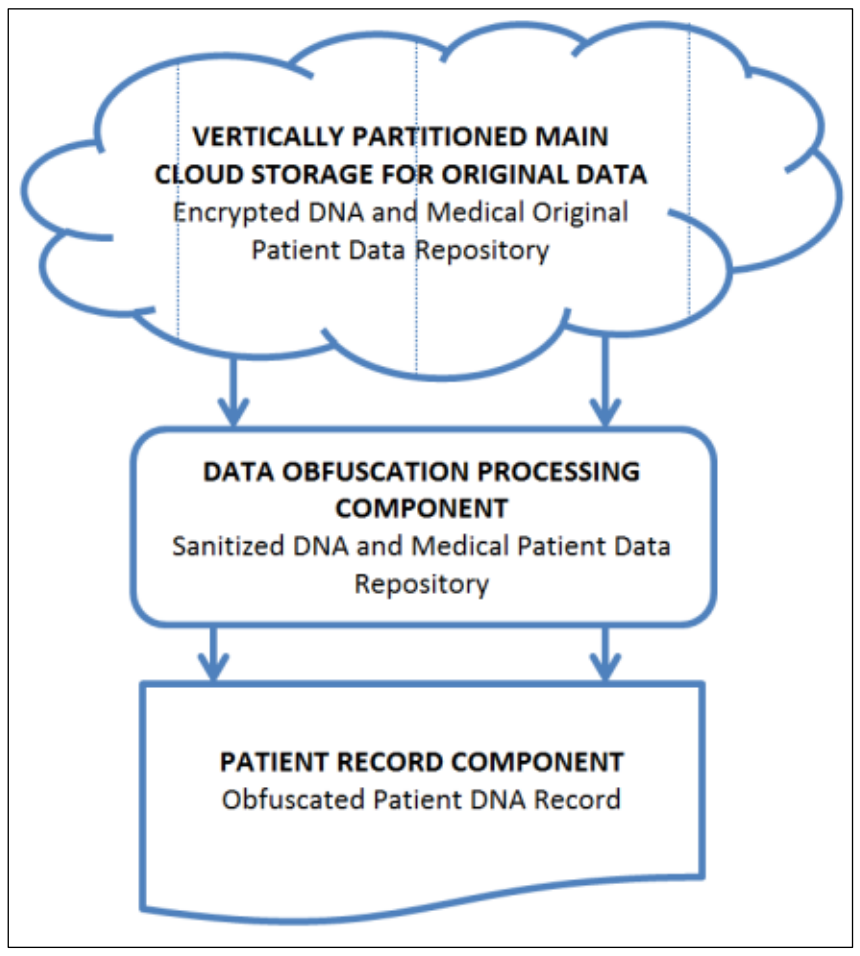

**Figure 2:** Overview of the bio-bank and DNA transaction model

From a bio-bank architectural privacy modeling view, we follow a similar approach to Ayday et al., (2014) by proposing the encryption of all medical records, including genomic data at the bio-bank level. However, in our model, the full perturbed raw DNA sequence, in which sensitive genomic information is either suppressed or altered, is retrieved. In our model, the privatized DNA sequences are retrieved by controlling the number of queries to prevent information leakages. For instance, a clinical researcher in hospital *A*, might be interested only in a set of genes responsible for cancer, while another researcher at hospital *B*, might be interested in genes linked to diabetes. In such scenarios, information leaked from hospitals *A* and *B* could be used to reconstruct sensitive information in the DNA sequence of a patient. Another way to get around the issue of information leakage, we propose a publication of static tabulated results for non-interactive settings. In this scenario, our model would only publish or transact in pre-defined results that could not be changeable, after the initial obfuscation. In this case, patients would publish the same privatized DNA sequence to all parties involved. This would be useful when dealing with entities such as, insurance companies and clinical researchers, only interested in aggregated statistical analysis.

## 4. Methodology

In this section we present the proposed methodology on the obfuscation of codons with highly expressed genes, for privacy. In this paper we assume that the DNA data transaction will occur between the patient, data curator (bio-bank), and a clinical researcher. The goal is to obfuscate the DNA data so that the clinical researchers will not view any gene information deemed sensitive by the patient. In other words, the patient controls by authorization, what they wish to be disclosed. Our proposed architecture, as illustrated in Figure 2, is composed of three components, the main original data storage component, the obfuscated data component, and the patient record component. *The main original data storage component (Bio-bank)*: This acts as a main storage for the patient medical records storage, including the DNA data. The original data storage component is highly restricted and all records are encrypted. Access to this data storage is enforced by documentation of who gained access, what time they gained access, how long they accessed the system, what information they access, and what time they exited the storage system. In this controlled environment, access to the storage requires another signatory for accountability. In other words, the system will not allow one individual to access the system without accountability from another official.

*The obfuscation data processing component*: This component, as illustrated in Figure 3, acts as a privacy layer between the patient record access component and the main original data storage component. Data generation in the obfuscated data storage is done by controlled query access on the main original data storage component. The number and type of queries run on the system is strictly limited to prevent information leakage. Obfuscation procedures are then applied to the results that are generated by the query. The only records stored or accessed in the obfuscated data storage component, are those that have been sanitized (privatized) via obfuscation and other data privacy algorithms.

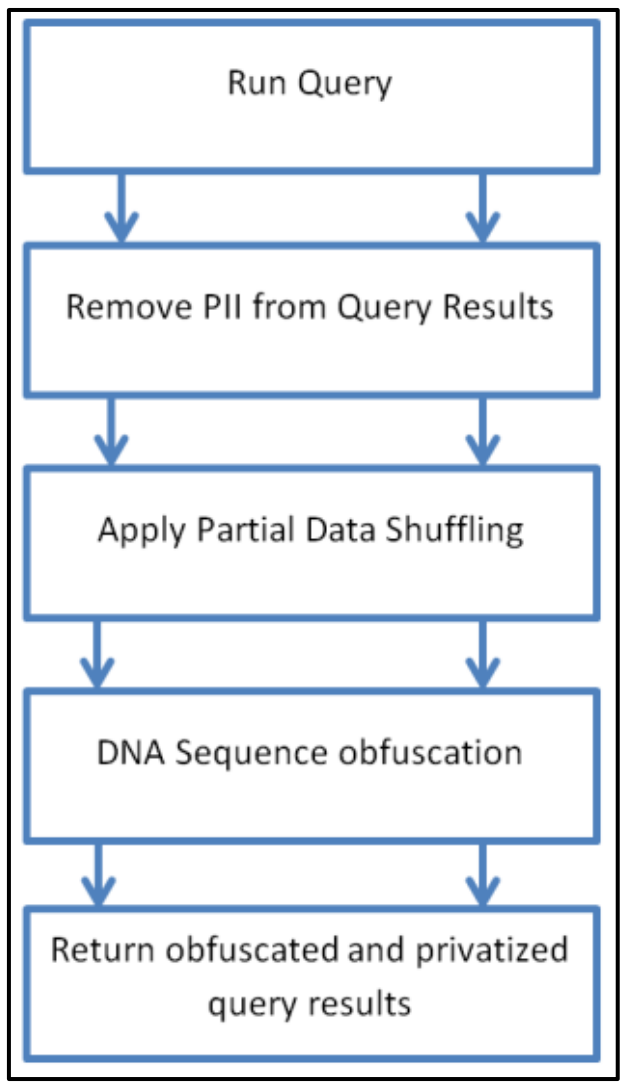

**Figure 3:** The obfuscated data processing component

Access to the obfuscated data component and querying process is strictly enforced using the same security access control methods, as applied to the original data storage component. Any data transaction, will only involve access to the Obfuscated data storage unit. This component is responsible for generating obfuscated patient records needed for any data transaction. *The patient record component*: In this component, the patient and or clinical researcher have access to the obfuscated medical and DNA record. The patient can authorize time restricted access to this record to their physician or a clinical researcher. The patient record component will document each access by a doctor, clinician, or the patient themselves. The patient record component could be accessible using mobile devices, via secure and encrypted connectivity.

*Phase I:* In the first phase of the obfuscation, as illustrated in Figure 3, Personal Identifiable Information (PII) is removed from the data set. After PII is removed, data shuffling is applied to the de-identified data set to increase obfuscation. In our methodology, partial data shuffling is applied on the data set. All attributes undergo data shuffling with exception of the Patient ID and DNA attributes. The reason for the data shuffling is that if the controlled and restricted data falls in the hands of an attacker, it would be difficult to reconstruct the full identify of a patient. Yet still, this would prevent inside attacks by those who have minimal access to the data system at various technical support levels. In our methodology, data shuffling is based on a random number generator. The rows of data values (tuples) in each attribute are aligned to a sequence of randomly generated numbers, which then get sorted. The seed for the random number generator is kept secret by the data curators (Bio-bank), to make it difficult for any reverse engineering. In this model, each attributed is shuffled vertically, separately, and finally, all attributes are combined to generate the complete partially shuffled data set.

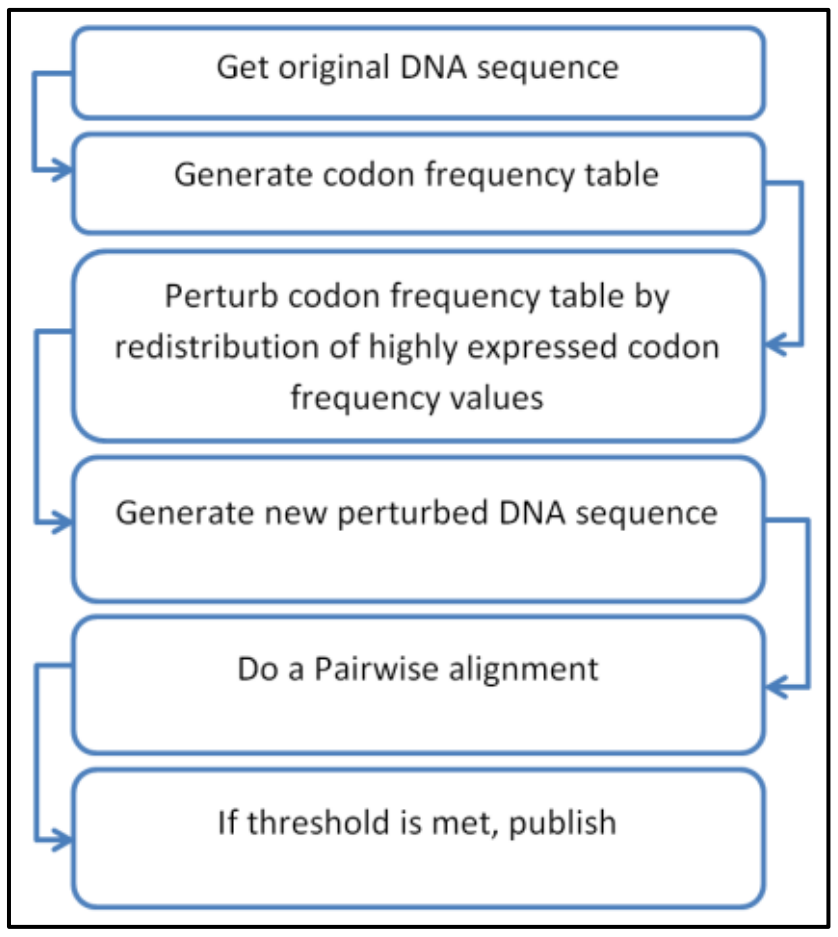

**Figure 4:** DNA data obfuscation overview

*Phase II:* In phase II, data obfuscation is done on the DNA sequence, by perturbation of the codon frequency table, as illustrated in Figure 4. In the first step, a codon frequency table is generated for each of the DNA sequence in the database. A frequency analysis is then done to highlight the most frequent codons in the table – that is, codons with the most highly expressed genes in each amino acid group. Perturbation by redistribution is done within the same amino acid group. Perturbation is done on codons with high frequency values – by redistributing the frequencies among the codon values in the same amino acid group. In this way, the same amino acid could remain 'highly expressed' while giving a form of obfuscation to the highly expressed genes in the codon.

**Table 1**, Codon frequency for Asp before obfuscation

| | | |
|---|---|---|
| Asp | GAT | 1 |
| Asp | GAC | 5 |

For example, the codon GAC has a high frequency value in the aspartic (Asp) amino acid group, with a frequency of 5 while GAT appears only once. The total frequency for the Asp amino acid group is then 6, as shown in Tables 1.

After perturbation by redistribution of the codon frequencies, the GAC codon gets a frequency value of 3 while the GAT codon, a frequency value of 3. However, the total frequency value for the Asp amino acid group remains intact, at 6, as show in Table 2.

**Table 2**: Codon frequency for Asp after obfuscation

| | | |
|---|---|---|
| Asp | GAT | 3 |
| Asp | GAC | 3 |

The totals of all codons in the frequency table stay the same; at this stage, the DNA sequence base pair size stays intact and the same as the original. After perturbation by redistribution of the codon frequencies, the new generated frequency table (privatized table) is then used to generate a new privatized DNA sequence. A pairwise alignment to compare the original DNA sequence to the privatized DNA sequence is done. Similarity values are used to determine if an acceptable threshold value is achieved. For example, a 50 percent similarity value might be acceptable as threshold; if this threshold value is not met, then a refinement is done by re-doing the redistribution of frequency values in the codon table. The last step is to publish the privatized DNA sequence without revealing data from highly expressed genes in codons with high frequencies. The original codon frequency values are kept encrypted at the main bio-bank storage. Therefore, this procedure could be applicable by implementing the extra layer of confidentiality for the patient DNA sequence during data transaction from one point to another.

## 5. Experiment and Results

*Experiment, Phase I:* To test our hypothesis, we generated a synthetic patient records data set with 150 data points containing PII information, medical records, and the DNA sequence of each patient.

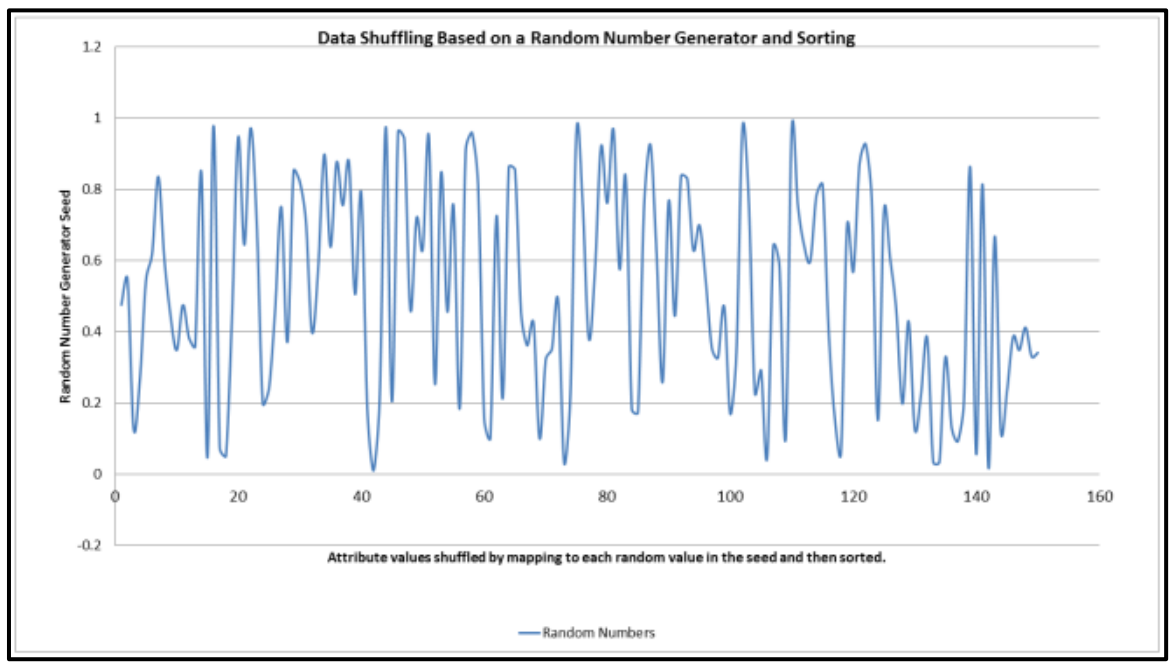

**Figure 5:** The random number generator seed used for shuffling

The synthetic patient records data set was used to simulate the real original patient records data set. The synthetic patient records included the following set of attributes: {Patient ID, First Name, Last Name,

Gender, SSN, DOB, Address, City, State, Zip code, Country, Diagnosis Code, Diagnosis Description, PCP, and DNA}. During the first phase of the obfuscation, the following attributes were removed from the data set to satisfy PII sanitization requirements: {First Name, Last Name, SSN, Address, and City}. Data shuffling was then individually applied to each of the following remaining attributes: {Patient ID, Gender, DOB, State, Zip code, Country, Diagnosis Code, Diagnosis Description, PCP, and DNA}.

| Patient_ID | Gender | DOB | State | Zipcode | Country | Diagnosis Code | Diagnosis Description | PCP | DNA |
|---|---|---|---|---|---|---|---|---|---|
| P000001 | M | 1979 | SD | 39967 | USA | 298.3 | Bronchitis, acute | Soo Mills | CAAGACAG |
| P000002 | F | 1991 | SD | 57037 | USA | 530.81 | Asthma, extrinsic, acute | Niverta Fulks | TACGACTC |
| P000003 | M | 1979 | NH | 17657 | USA | 300.3 | Atelectasis | Ruby Banh | ATCGATTG |
| P000004 | M | 1987 | NV | 04456 | USA | 074.3 | Anemia, other, unspec. | Bjorn Michala | TCTTTGAA |
| P000005 | F | 1977 | WA | 15278 | USA | 782.8 | Pain, psychogenic | Routledge Gu | TAGTATC |
| P000006 | F | 1979 | NH | 52217 | USA | 493.12 | Asthma, cough variant | Elizabeth Rug | CCGTGTAT |

**Figure 6:** Values in red show the change in position after shuffling

A random number generator was used in shuffling the values in each separate attribute. The values were mapped to the randomly generated numbers between a chosen seed, in the case of our experiment, the seed was chosen as real numbers between 0 and 1. After the mapping the values in the attributes to the random numbers, sorting was done. The values in the attributes changed positions with each sorting iteration, thus shuffling, as illustrated in Figure 6. Figure 6, shows an illustration of a returned query results showing the effects of shuffling on the attribute values. The highlighted figures in red in Figure 6, show the shift in each of the shuffled values from their original positions.

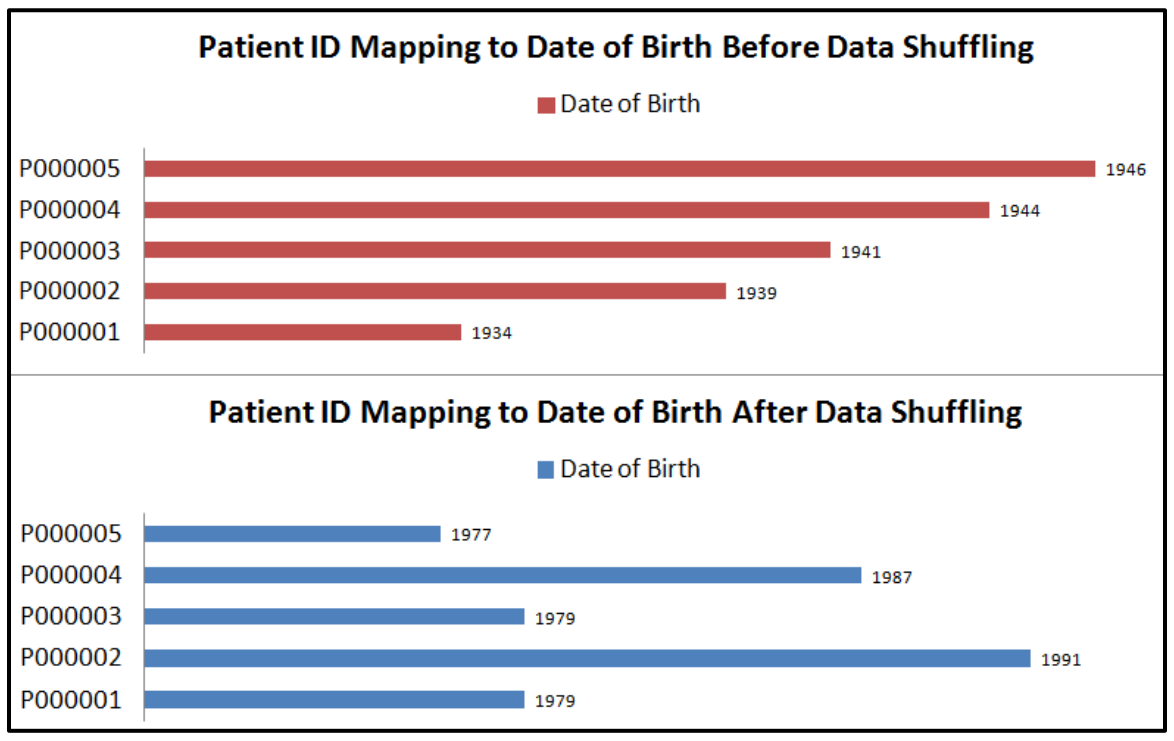

**Figure 7:** Shuffling results for the date of birth attribute.

It would be difficult for an attacker to reverse engineer the process to gain the original row of data without prior knowledge, and considering that the database would contain millions of data points. To ensure that the data values in each attribute were efficiently shuffled, we did a frequency distribution analysis on the date of birth attribute, before and after shuffling. As illustrated in Figures 7, the date of birth for Patient ID P000001 was 1934 before shuffling. However as illustrated in Figure 7, the date of birth for the same Patient ID P000001 was substituted with 1979 after shuffling, thus offering a layer of confidentiality for Patient ID P000001.

*Experiment, Phase II:* To further test our hypothesis, we generated synthetic DNA sequences for each of the Patient ID records in that database. For the remainder of this paper and context of this experiment, we shall refer to both the synthetic DNA sequence and synthetic patient records, as original data. Each generated original DNA sequence contained 500 base pairs. The FASTA format was used in formatting the raw generated original DNA sequence. 150 original DNA sequences were generated for the original patient database. In the next step of the Phase II experiment, the codon frequency of the generated original DNA sequence was done. Codon frequency obfuscation was achieved by redistributing values of codons with highly expressed genes, within the same amino acid group. For instance as illustrated in Figure 8, the codon values for CAG and CAA in the Glutamine (Gln) amino acid group was 2 and 9 respectively. In the obfuscation process, the codon values for CAG and CAA were reassigned to 5 and 6 respectively, after redistribution.

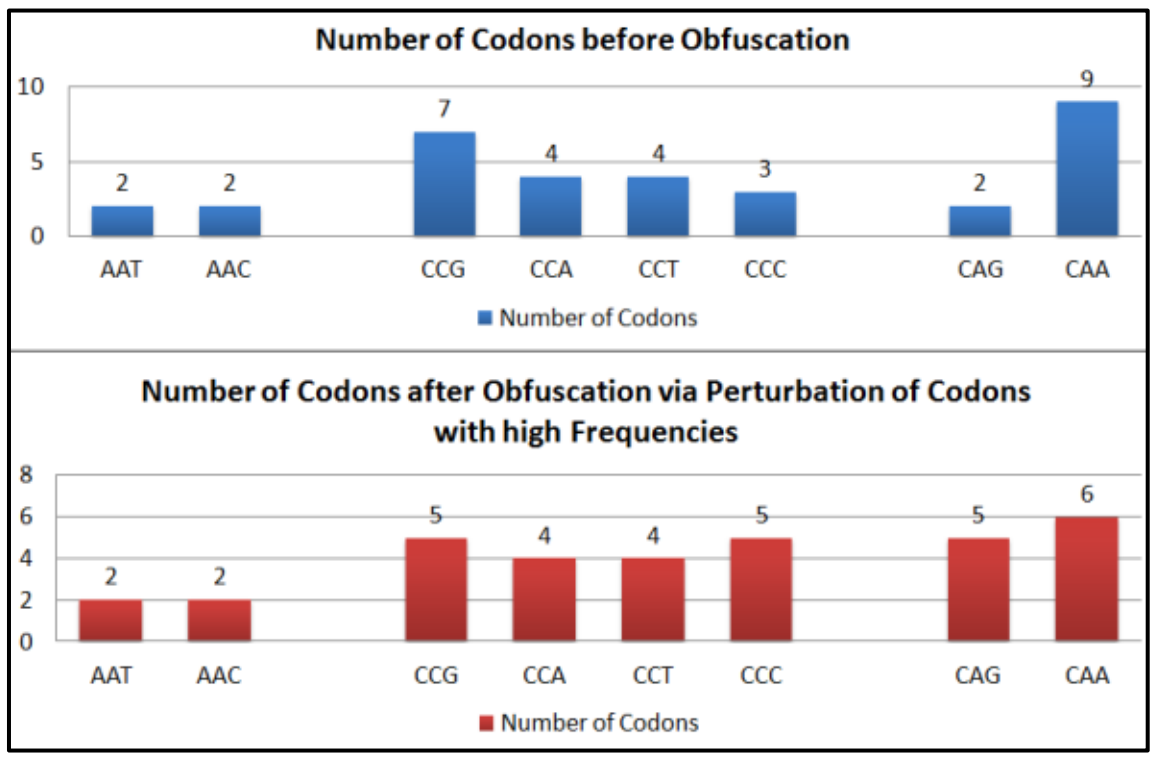

**Figure 8:** Codon frequencies before and after obfuscation.

It is important to note that while the codon values for CAG and CAA were altered, the totals for the frequency of that particular amino acid, in this case Glutamine (Gln), remained the same at 11. In other words, perturbation was done to the frequency values of the codons themselves but not the frequency values of the amino acid group. As shown in Figure 9, the frequency of the Glutamine amino acid in the original DNA sequence is 11, and after obfuscation, the frequency value for the same amino acid stays the same at 11. The obfuscated codon table was then used to generate the obfuscated DNA sequence. A pairwise comparison was then done on both the original and obfuscated DNA sequences. A pairwise analysis was done on both the original and obfuscated DNA sequence. The goal was to implement an obfuscation of codons with highly expressed genes in the DNA sequence. The question of data privacy versus usability still affects the outcome of the results. Too much obfuscation may alter the DNA sequence with many mutations; minimal obfuscation could also be revealing.

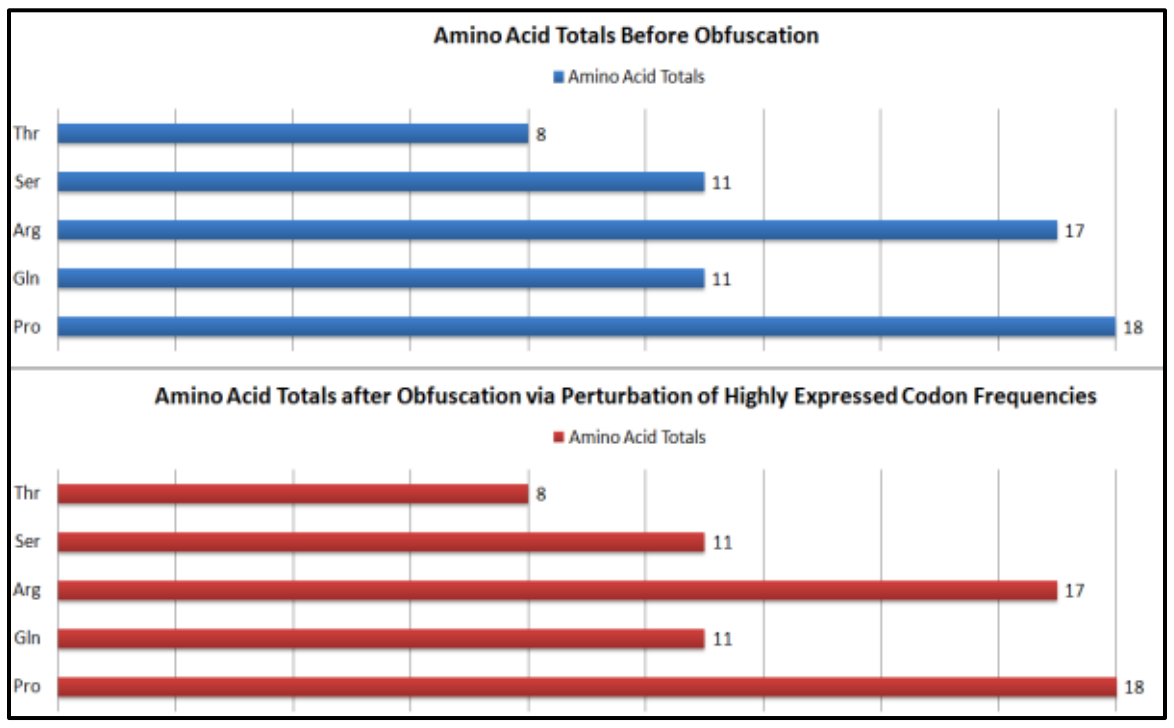

**Figure 9:** Amino acid frequencies before and after obfuscation.

In the first part of our pairwise analysis, we compared the original protein sequence to the obfuscated protein sequence. Tables 3 and 4 show the protein sequences we used for the pairwise analysis for both the original and obfuscated data.

Table 3: Original protein sequence

```
  >EMBOSS_001_1
  QDSSGLEVAPAKEDAWYSRTMKPVQRQHP
HLGEPKRGIKVTFWIPETNPMVVLRLGSLSP
  LGPISGNRRQVQSVFRRTMEQCG*DVLHL
YPTQAGPRSARRGGAQAIDN*PPCIHYGTRD
  FKPSQWSSQYRVYRILP*LTNCDPPQVKPLP
LRHAVRVIM*TLRGLT

 Sequence Translation tool:
http://www.ebi.ac.uk/Tools/st/
```

Table 4: Obfuscated protein sequence

```
  >EMBOSS_001_1
  RDAQSVPGSSTVERVLGTHDTADDVIYRT
SHLASLGRRDQYSTNEPNSSRPRSLPRTGAA
  EPLLGGLQPRHYSRHPLRHGAWRKAGRNY
KRNAQVVAFNQLSLSVRSSQPFKLSRHST*Q
  LFASGTGLSD*WQPHVLL*LP**LPGPH*L
GSRYRHGPCFNNQIF

 Sequence Translation tool :
http://www.ebi.ac.uk/Tools/st/
```

As shown in Figure 10 and 11, the pairwise alignment analysis for both the original and obfuscated protein sequence, generated only 18 percent similarity [20]. While this might be a good indication of privacy and obfuscation, questions remain as to the viability of DNA data utility. The dot matrix in Figure 11 shows regions of similarity between the original and obfuscated protein sequences. The *x*-axis represents the obfuscated protein sequence while the *y*-axis shows the original protein sequence. A diagonal line in the matrix would represent an ideal perfect match for similarity. The nonaligned lines represent the mismatch between the two protein sequences.

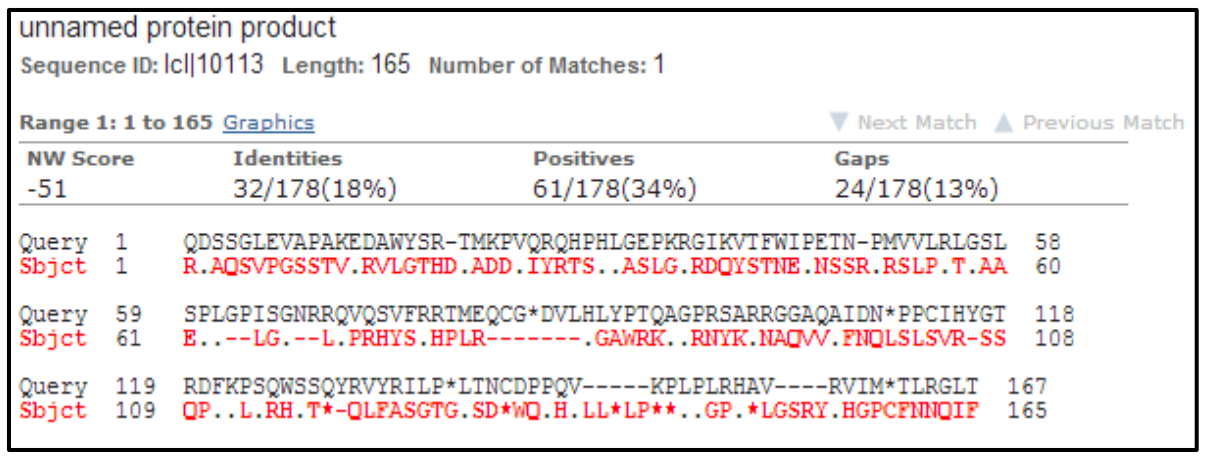

```
unnamed protein product
Sequence ID: lcl|10113  Length: 165  Number of Matches: 1

Range 1: 1 to 165 Graphics                         Next Match  Previous Match
NW Score     Identities       Positives       Gaps
-51          32/178(18%)      61/178(34%)     24/178(13%)

Query  1    QDSSGLEVAPAKEDAWYSR-TMKPVQRQHPHLGEPKRGIKVTFWIPETN-PMVVLRLGSL  58
Sbjct  1    R.AQSVPGSSTV.RVLGTHD.ADD.IYRTS..ASLG.RDQYSTNE.NSSR.RSLP.T.AA  60

Query  59   SPLGPISGNRRQVQSVFRRTMEQCG*DVLHLYPTQAGPRSARRGGAQAIDN*PPCIHYGT  118
Sbjct  61   E..--LG.--L.PRHYS.HPLR-------.GAWRK..RNYK.NAQVV.FNQLSLSVR-SS  108

Query  119  RDFKPSQWSSQYRVYRILP*LTNCDPPQV-----KPLPLRHAV----RVIM*TLRGLT  167
Sbjct  109  QP..L.RH.T*-QLFASGTG.SD*WQ.H.LL*LP**..GP.*LGSRY.HGPCFNNQIF  165
```

**Figure 10:** The pairwise alignment between original and obfuscated protein sequences.

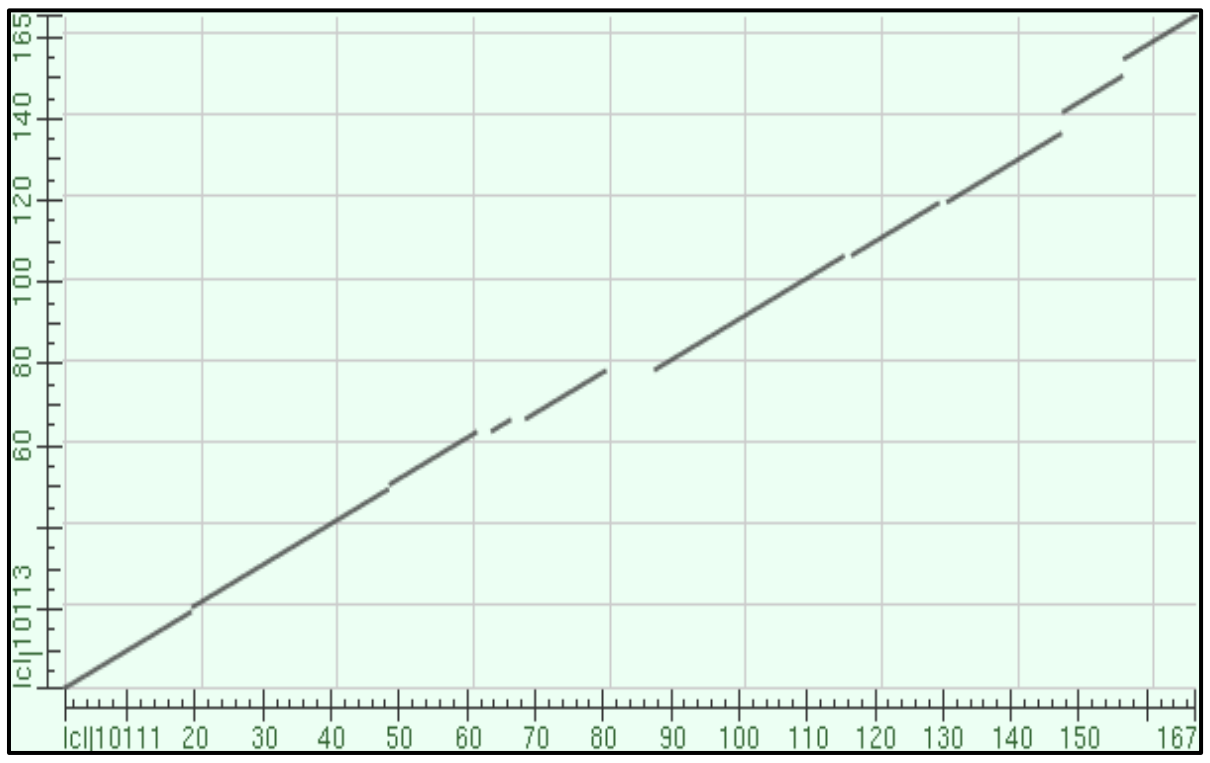

**Figure 11:** The original and obfuscated protein sequence dot matrix.

We then used the protein sequence of the original DNA sequence to act as a key and reverse engineered the obfuscated DNA sequence to obtain the original DNA sequence by reverse translation.

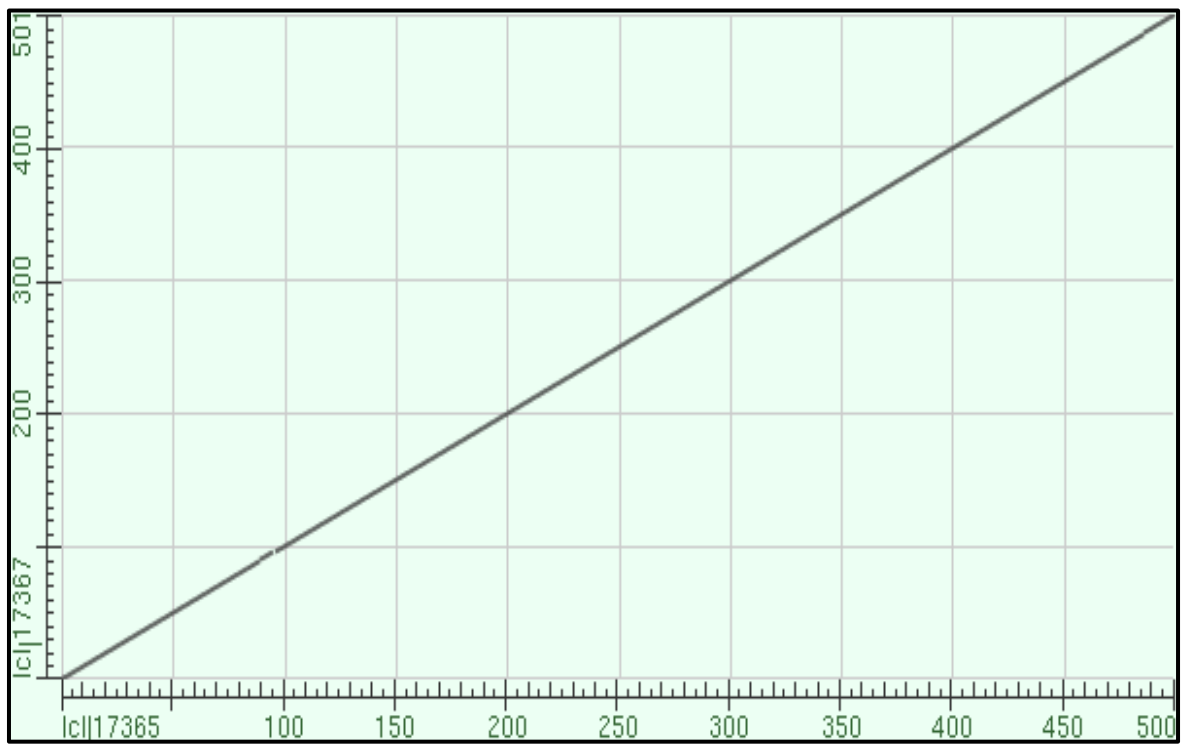

**Figure 12:** The original and obfuscated DNA sequence dot matrix, after reverse translation.

We used the codon frequency table of the obfuscated DNA sequence to generate the original DNA

sequence, using the protein sequence of the original DNA sequence as key. The dot matrix in Figure 12 shows regions of similarity between the original and obfuscated DNA sequences after the reverse engineering process. The *x*-axis represents the obfuscated DNA sequence while the y-axis shows the original DNA sequence. A diagonal line in the matrix would represent an ideal perfect match or similarity. The nonaligned lines represent the mismatch between the two DNA sequences. As shown in Figure 12 and 13, there was a 78 percent similarity between the original and obfuscated DNA sequences after the reverse engineering process.

```
Sequence ID: lcl|17367  Length: 501  Number of Matches: 1

Range 1: 1 to 501 Graphics                          ▼ Next Match  ▲ Previous Match
NW Score    Identities       Gaps         Strand
442         392/502(78%)     3/502(0%)    Plus/Plus

Query  1    CAAGACAGTTCCGGGCTGGAAGTAGCGCCGGCTAAGGAAGACGCCTGGTACAGCAGGACT  60
Sbjct  1    .....T...AGT.....T..............G........T..G........T......  60

Query  61   ATGAAACCGGTACAAAGGCAACATCCTCA-CTTGGGTGAACCGAAACGCGGTATCAAGGT  119
Sbjct  61   .....G....................G..T.....-........GA.G..G..T.....  119

Query  120  TACTTTTTGGATACCTGAAACAAATCCCATGGTAGTCCTTAGACTTGGGAGTCTATCACC  179
Sbjct  120  A...........T..G.....T.....G........A.....G...........TAGT..  179

Query  180  CCTAGGGCCCATATCTGGAAATAGACGCCAAGTTCAATCCGTATTCCGACGTACGATGGA  239
Sbjct  180  G..T.....G..TAG...G.....GA.G.....A...AGT.....TA.GA.G..T.....  239

Query  240  ACAGTGTGGGTGAGACGTGCTTCATTTATACCCTACGCAGGCTGGACCGAGGTCCGCAAG  299
Sbjct  240  ...A.......A...T..A......C.T.....G..T..A..G..G......AGT..G..  299

Query  300  GCGCGGCGGTGCACAAGCAATTGACAACTAACCACCGTGTATTCATTATGGTACCAGGGA  359
Sbjct  300  .A.G..G..G..G.....G.....T..T.....G..............C..G..T.....  359

Query  360  CTTTAAGCCGAGTCAATGGAGCTCGCAATACAGAGTTTACCGCATCTTGCCGTAACTGAC  419
Sbjct  360  T...................TAGT........G..A...A.G..TC.T........T..  419

Query  420  AAACTGTGATCCACCACAAGTCAAGCCATTGCCTCTTAGACACGCCGTTAGAGTAATTAT  479
Sbjct  420  T..T........G..G.....A.....GC.T..G.....G..T..G..A..G........  479

Query  480  GTAAACT-TTGCGCGGCTTGAC  500
Sbjct  480  .......C..AG.G.....ACT  501
```

**Figure 13:** The dissimilarities between the original and obfuscated DNA sequences after reverse translation.

While the focus of this paper was not to study various reverse engineering attacks, our results show that when the protein sequence of the original data was used in combination with the codon frequency table of the obfuscated DNA data, to reserve translate, there was only a 78 percent success rate.

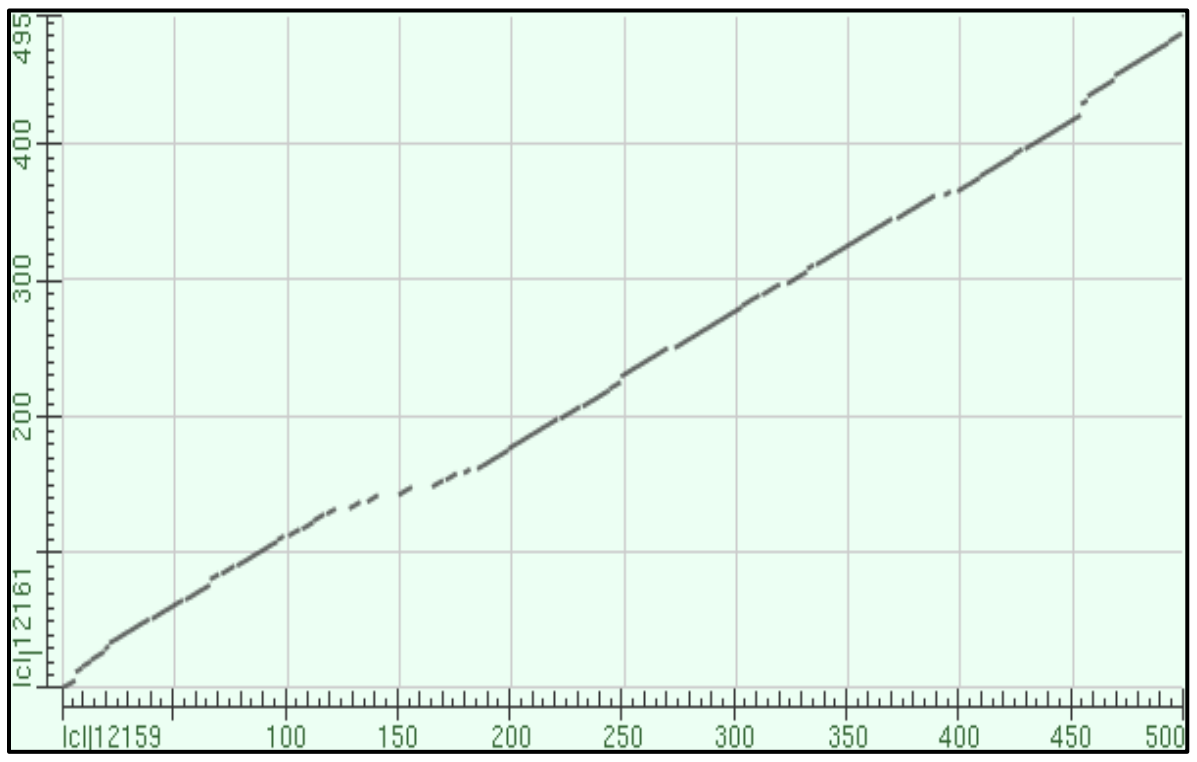

**Figure 14:** The original and obfuscated DNA sequence dot matrix.

The results could as well mean that this heuristic was good for privacy, with 22 percent of the data remaining concealed – relating to the obfuscation done only on codons with high frequencies and highly expressed genes. In Figure 13, the highlighted obfuscated DNA sequence in red, represent the dissimilarities between the original and obfuscated DNA sequences after the reverse translation process. The dot matrix in Figure 14 shows regions of similarity between the original and obfuscated DNA sequences. The *x*-axis represents the obfuscated DNA sequence while the *y*-axis shows the original DNA sequence. A diagonal line in the matrix would correspond to a perfect similarity. The nonaligned lines represent the mismatch between the two DNA sequences. The results in Figure 14 and 15 show the pairwise alignment between the original DNA sequence and the obfuscated DNA sequence generated from the perturbed codon frequency table without any the reverse engineering process. The results show that there was a 50 percent similarity, indicating that privacy might be achieved by the concealment provided by the other 50 percent dissimilarity; yet again, the issue of privacy versus usability resurfaces.

```
Sequence ID: lcl|12161  Length: 495  Number of Matches: 1

Range 1: 1 to 495 Graphics                          ▼ Next Match  ▲ Previous Match
NW Score    Identities       Gaps            Strand
-392        276/557(50%)     119/557(21%)    Plus/Plus

Query  1    CAAGAC------AGT-TCCG-GGCTGG--AA--GTAGCGCCGGCTAAGGAAGACGCCTGG  48
Sbjct  1    .G....GCACAA...G...CT..A.C.TC..CG..G.A..GT.T.TT..GC-....A..A  59

Query  49   TACAGCAGGACTATGAAA----CCGGTACAAAGGCAACATCCTCACTTGGGT-GAACCGA  103
Sbjct  60   ......-...TG...T..TCTA....-...TC.-..C.TGG..AG...A...C.G.--..  114

Query  104  AACGCGGTA-TCAAGGTTACTTTTTGGATACCTGAAACAAATCCCATGGTAGTCCTTAGA  162
Sbjct  115  G..-.A...C....C-.A..------..GC.---.....---------....C.------  148

Query  163  CTTGGGAGTCTATCACCCCTAGGGCCCATATCTGGAA-ATAGACGCCAAGTTCAATCCGT  221
Sbjct  149  ---..CC.-.G...---...---.....G.A.G...GC.GC.GAA..GCT..TGGGGG..  198

Query  222  ATTCCGACGTACGATGGAACAGT-GTGGG----TGAGACGTGCTTCATTTATACCCTACG  276
Sbjct  199  -..A.A..C.-...CACT.....C..CATCCTC......A..GGG.T.GG.G.A---.A.  253

Query  277  CAGGCTGGACCGAGGTCCGCAAGGCGC-GGCGGTGCACAAGCAATTGACAACTAACC--A  333
Sbjct  254  ....GC.T.A.T.CAAG.....T..T.A..TA...G-.TTT....---.......G.TT.  309

Query  334  CCGTGTATTCATTATGGTACCAGGGACTTTAAGCCGAGTCAATGGAGCTCGCAATACAGA  393
Sbjct  310  T.-...GAGA.GCTC.CAG...TTT.AA...TC...G--..T.C..CG.A.....T----  362

Query  394  GTTTACCGCATCTTGC-CGTAACTGACAAACT-GTGATCCACCACAAGTCAAGCCATTGC  451
Sbjct  363  ...---...C..AG..A..GGGT..T.GG.T.A...G-.A......T..ATTA.TC.A..  418

Query  452  CTC--------TTA---GACACGCCGTTA---GAGTAATTATGTAAACTTTGCGCG-GCT  496
Sbjct  419  T..CATAATGA...CCT.G.C.A.AC.A.CTG.G..C.CG..A.CGT.ACG..C.CT...  478

Query  497  TGA------------C  500
Sbjct  479  .C.ATAACCAGATTTT.  495
```

**Figure 15:** The dissimilarities between the original and obfuscated DNA sequences.

## 6. Conclusion

We have presented a codon frequency obfuscation heuristic for genomic data privacy. DNA data privacy in general, remains a challenge. However, data usability remains a challenge with DNA obfuscation – more mutations in obfuscated data. Our proposed model shows that it might be possible to perturb the codon frequency table by redistributing codon frequencies of highly expressed genes within the same amino acid group for confidentiality. Our preliminary results show that even with reverse engineering using the obfuscated DNA codon frequency table in combination with the protein sequence of the original DNA sequence, only 78 percent was recoverable, indicating that it might be possible to conceal certain sections of information within the DNA sequence. In this proposed heuristic, we envision a well informed patient with full rights and authorization as to what information within the DNA sequence they might want to remain concealed. While the focus in this paper was to present preliminary results from the testing of the hypothesis, future works will include testing the model against various reverse engineering attacks. For future works include, running the same tests using real DNA data from living organism with large base pairs (big data). We plan on employing other data privacy algorithms not covered in this paper; developing a prototype of the obfuscation architecture to automate the proposed heuristic.